\documentclass[conference,compsoc]{IEEEtran}
\usepackage{graphicx}
\usepackage{amsmath,amssymb}
\usepackage{psfrag}
\usepackage{color}
\ifCLASSOPTIONcompsoc
  \usepackage[nocompress]{cite}
\else
  \usepackage{cite}
\fi
\ifCLASSINFOpdf
\else
\fi
\begin{document}
\pagenumbering{Alph}
\pagestyle{empty}
\title{Continuous and Discontinuous Phase Transitions in the evolution of a
polygenic trait under stabilizing selective pressure}
\author{
\IEEEauthorblockN{Annalisa Fierro}
\IEEEauthorblockA{
CNR-SPIN Complesso Univ. Monte S. Angelo, Via Cinthia, I-80126 Naples, Italy\\
Corresponding Author - Email: annalisa.fierro@spin.cnr.it}
\and
\IEEEauthorblockN{Sergio Cocozza}
\IEEEauthorblockA{
Dipartimento di Medicina Molecolare e Biotecnologie Mediche, \\
Universit\`a degli Studi di Napoli ``Federico II", Naples, Italy}
\and
\IEEEauthorblockN{Antonella Monticelli}
\IEEEauthorblockA{
Istituto di Endocrinologia ed Oncologia Sperimentale, CNR Napoli, \\
Naples, Italy}
\and
\IEEEauthorblockN{Giovanni Scala}
\IEEEauthorblockA{INFN - Sez. Napoli, Complesso Univ. Monte S. Angelo, \\
Via Cinthia, I-80126 Naples, Italy}
\and
\IEEEauthorblockN{Gennaro Miele}
\IEEEauthorblockA{
Dipartimento di Fisica ``Ettore Pancini'', Universit\`a degli Studi di Napoli ``Federico II",\\ 
and INFN - Sez. Napoli, Complesso Univ. Monte S. Angelo, \\
Via Cinthia, I-80126 Naples, Italy}}
\maketitle

\begin{abstract}
The presence of phenomena analogous to phase transition in Statistical Mechanics, has been suggested in the evolution of a polygenic trait under stabilizing selection, mutation and genetic drift.

By using numerical simulations of a model system, we analyze the evolution of a population of $N$ diploid hermaphrodites
in random mating regime. The population evolves under the effect of drift, selective pressure in form of viability on an
additive  polygenic trait, and mutation.  The analysis allows to determine a phase diagram in the plane of mutation rate
and strength of selection.  The involved pattern of phase transitions is characterized by a line of critical points for weak selective
pressure (smaller than a threshold), whereas discontinuous phase transitions, characterized by metastable
hysteresis, are observed for strong selective pressure.

A finite size scaling analysis suggests the analogy between our
system and the mean field Ising model for selective pressure approaching the threshold from weaker values.
In this framework, the mutation rate, which allows the system to explore the accessible microscopic states, is the
parameter controlling the transition from large heterozygosity ({\em disordered} phase) to small heterozygosity ({\em ordered} one).

\end{abstract}


%
\IEEEpeerreviewmaketitle

\section{Introduction}
The need of applying a statistical approach arises in physics when the global properties of ensembles containing a huge number of elementary constituents are studied. In this case, the behavior of the single element is irrelevant in favor of more informative averages on the whole ensemble.
In this sense, the situation is strictly analogous in both Population and Quantitative Genetics, where the focus is on the analysis of allele frequencies and phenotype distribution parameters rather than the genetic/phenotypic description of each individual. The analogy between Quantitative Genetics and thermodynamics  was noted from the very beginning  by R. A. Fisher himself \cite{Fisher1930}.
The development of such analogy allowed Iwasa to introduce a concept of entropy for Population Genetics, which satisfies the analogous of H-theorem \cite{Iwasa1988}. Such information entropy measure ensures an exact solution at statistical equilibrium  \cite{Iwasa1988,Barton1989,Sella2005,Barton2009}. This point of view, can be alternatively seen by defining a free fitness, namely the entropy divided by population size plus the mean fitness.
The free fitness is maximized at equilibrium, when natural selection and drift (random sampling) are at work \cite{Iwasa1988,Barton2009}, and provides an analogous of free energy in thermodynamics.

Many analogies between biological evolution and statistical physics are present in literature (for a review see \cite{baakegabriel,drossel}, and
reference therein). The presence of phenomena analogous to phase transition has been also suggested
\cite{BaakeWagner,ParkKrug,deVladar2011}. 

In this paper, we analyze in details such phase transition phenomena in a stochastic model, extensively analyzed in the past years (see for instance Ref. \cite{Barton1989}, and references therein). It consists in a population of $N$ diploid hermaphrodite individuals, reproducing in pairs in a random mating regime, evolving under the effect of drift, selective pressure in form of viability, and mutation (we assume ``substitution'' mutations). Following Wright's seminal paper \cite{Wright:1931}, we consider $M$ different bi-allelic genes additively combining on the character, and the individual viability following a Gaussian profile in the trait.
Using numerical simulations of such a model, we determine a phase diagram in the 
plane of mutation rate and strength of selection.
The involved pattern of phase transitions is characterized by a transition from a state, where the alleles of individuals are roughly
randomly distributed, to a state of clones, where individuals display a unique genome. This transition presents feature of a second order
transition for weak selective pressure
(smaller than a threshold), whereas discontinuous phase transitions, characterized by metastable
hysteresis, are observed for strong selective pressure.

\section{The model}
Using numerical simulations, we study a model for $N$ diploid individuals, sexually reproducing with random mating between any pairs of individuals.  Each individual $i$ (with $i=1, \dots, N$) is represented by two sequences of $M$ variables, $\sigma_{ikj}$ (where $k=1,2$ stands for the two genome replicas, and $j=1,\dots,M$ runs on  different loci). We refer to $\sigma_{ikj}$ as alleles, and assume that each allele can take two values, $\pm 1$.  A mutation rate, $\mu$, is introduced, as the probability of an allele to mutate at each generation ($\sigma_{ikj} \rightarrow -\sigma_{ikj}$). Similar models were studied in Refs. \cite{ServaPeliti1991, HiggsDerrida1991}.

A stabilizing selective pressure on such a phenotype can be  implemented via the survival probability of an individual $i$, typically known as {\it viability},
\begin{equation}
S(p_i)\equiv{\cal N}  \exp\left[-(p_i-p_m)^2 \omega^2/2\right],
\label{viability}
\end{equation}
where ${\cal N}$ is the suitable normalization constant,  $\omega$ measures the {\it strength of selection}, $p_i=\sum_{j=1}^M \sum_{k=1}^2  \sigma_{ikj}$ is the additive polygenic phenotype random variable,
and $p_m$ stands for the optimum phenotype.

For $\omega\rightarrow0$, there is no selective pressure, and all the microscopic states are equivalent.  Whereas, for $\omega\rightarrow \infty$,  the selective pressure is at maximum, and the only surviving individuals are those with $p_i=p_m$. In general, the effect of the selective pressure is to reduce the accessible phase space to those microscopic states better conform to the constraint on the phenotype.

Numerical simulations of the model are performed for different set of the parameters $N$, $M$, $\omega$, and $\mu$.
Starting from a common initial state, where the $M$ variants are chosen equal to $\pm 1$ with  equal probability, $30$ independent populations evolve with different random noise.
Our choice of the initial state corresponds to an initial frequency of the allele  ``$1$'' in locus $j$,  $\rho_{\rm in}(j)=0.5$, $\forall j=1, ..,M$.
During the evolution
\begin{enumerate}
\item two individuals, $i_1$ and $i_2$, are randomly
chosen and an off-spring $i$ is generated, such that
$\sigma_{ikj}$ of the off-spring is equal to $\sigma_{i_1 kj}$ or
$\sigma_{i_2 kj}$ with equal probability;
\item
the alleles are mutated (i.e., $\sigma_{ikj} \rightarrow -\sigma_{ikj}$)
with probability $\mu$ (called mutation rate);
\item
the newborn individual survives with probability $S(p_i)$, given by Eq. (\ref{viability});
\item
the point 1-3 are iterated until $N$ newborn individuals are generated. Then, the old generation is replaced by a new generation of same size $N$, formed by off-springs of the previous individuals. Note that the population size is fixed and not allowed to fluctuate.
\end{enumerate}
The equilibrium results are independent of the initial assumption about $\rho_{\rm in}(j)$.
Further analysis is necessary to evaluate the effect of the initial state on the out-of-equilibrium behavior. Hereafter, we choose $p_m =0$, however preliminary simulations show the model with a different optimum ($p_m \ne 0$) displays
qualitatively similar behavior.

The connection between the present model and a usual system of Statistical
Mechanics with Ising spins is rather natural (one can speculate that the
random mating is similar to a long range interaction between pairs of spins
in the same locus). The biological model in absence of selective pressure
remembers $M$ independent systems of $2N$ spins, the introduction of a
selective pressure instead corresponding to a coupling between different
systems. With this analogy in mind, we introduce the following quantities in
order to describe the macroscopic state of the biological model:
\begin{equation}
\overline{q}\equiv\frac{1}{M}\sum_{j=1}^M \langle |q(j)|\rangle,
\label{magn}
\end{equation}
with
\begin{equation}
q(j)\equiv\frac{1}{2N}\sum_{i=1}^{N} \sum_{k=1}^2\sigma_{ikj},
\label{magnj}
\end{equation}
where $\langle \dots \rangle$ stands for the average over the independently evolving populations (hereafter simply denoted by ensemble of populations). In our Statistical Mechanics analogue, the quantity $q(j)$, Eq. (\ref{magnj}), should correspond to the magnetization per spin in a system of $2N$ Ising spins, and  $\overline{q}$, Eq. (\ref{magn}), to the average of the magnetization modulus over different systems. Following the same analogy, we also introduce the susceptibility,  as
\begin{equation}
\overline{\chi}\equiv\frac{1}{M}\sum_{j=1}^M \chi(j),
\label{susc}
\end{equation}
with $ \chi(j)\equiv  2N \left( \langle q(j)^2\rangle-\langle |q(j)|\rangle^2 \right)$.
It is interesting to note that the magnetization $q(j)$ of j-th locus is
related to the expected heterozygosity (fraction of heterozygous individuals
expected on the basis of Hardy-Weinberg equilibrium condition) in the same
locus, denoted by $h_s(j)$, which is a more familiar quantity in the
Population Genetics context. Indeed, one can easily prove that
\begin{eqnarray}
h_s(j)&\equiv&\frac{1}{4N^2}\sum_{i,l=1}^{N} \sum_{k,n=1}^{2} 
\left(1-\delta_{\sigma_{ikj}\sigma_{lnj}}\right)=\nonumber\\
&=&\frac{1}{2}\left(1-q(j)^2\right).
\label{het}
\end{eqnarray}
In our case, $h_s(j)$ essentially coincides with the observed heterozygosity (observed fraction of heterozygous individuals).  From Eq. (\ref{het}), we see that the minimum of magnetization corresponds to the maximum of heterozygosity, and vice-versa. Denoting with $\overline{H}_s$ the average of $h_s(j)$ over different loci and on the ensemble, one can easily prove that
\begin{equation} 
\overline{H}_s = \frac 12 \left(1 - \frac{\overline{\chi}}{2N} - \overline{q}^2 \right)
\label{HS}
\end{equation}
for a very large number of realizations. This occurs since, in this limit,
$\langle |q(j)|\rangle$ is independent of $j$.

\section{Results and Discussion}
For any fixed set of the parameters, we follow the evolution of a given population till it asymptotically reaches a stationary state, which we refer to as {\em steady} state, where we evaluate $\overline{q}$ and $\overline{\chi}$. Let us start by focusing our attention on the role played by mutation rate and selection strength only, and to this aim we fix the values of $N=1000$ and $M=50$.  In general, the system reaches the steady state for values of generation number, which depend on $\omega$ and $\mu$. Two different behaviors are observed in the regime of small and large selective pressure strength, respectively.
\begin{figure}[ht]
\begin{center}
\includegraphics[width=6cm]{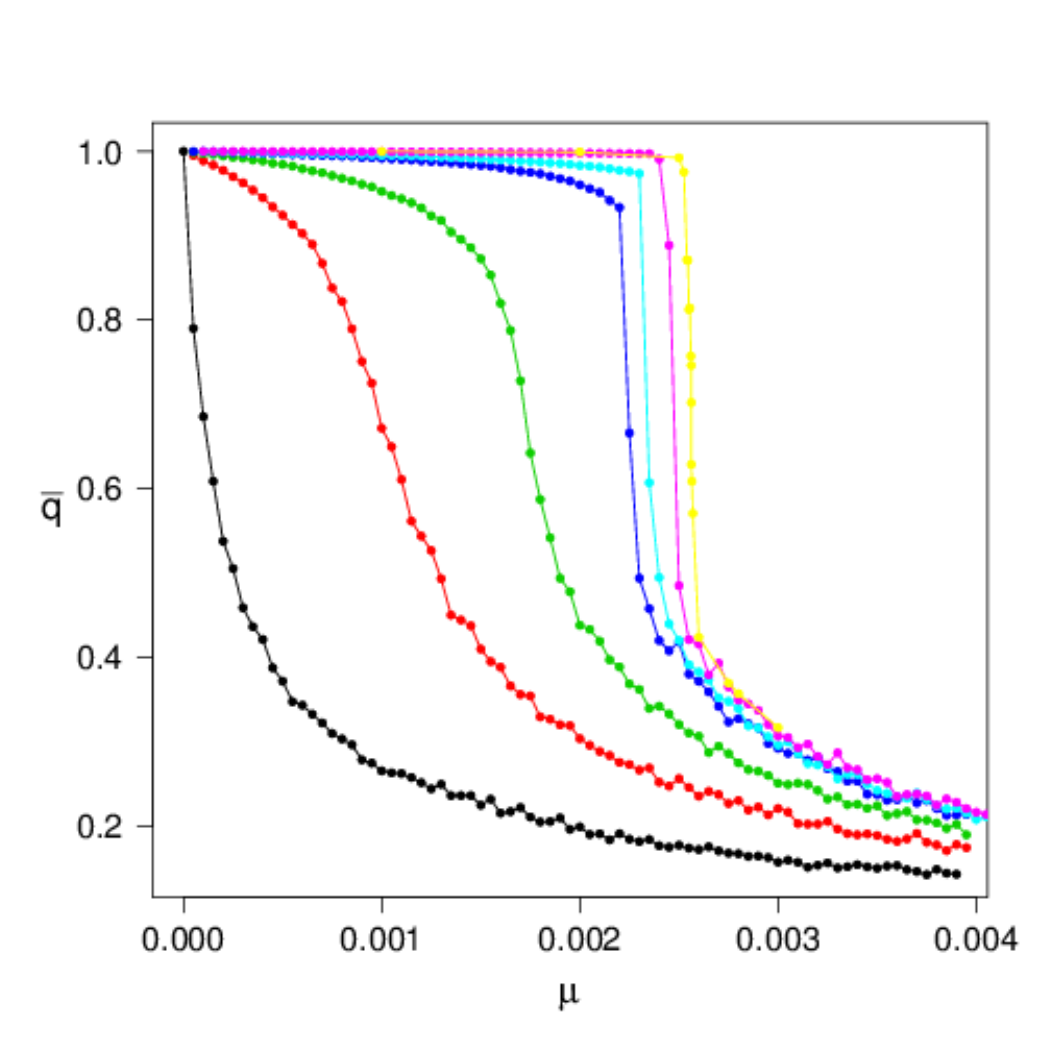}
\end{center}
\caption{Order parameter, $\overline{q}$ vs $\mu$, in the steady states for $N=1000$ and $\omega=0,~0.1,~0.2,~0.4,~0.5,~1,~\infty$ (from left to right). The continuous lines are guides for eyes.}
\label{fig:1}
\end{figure}

In Fig. \ref{fig:1}, we report $\overline{q}$ as a function of $\mu$ for different values of $\omega$. As shown in figure,  by decreasing the mutation rate $\mu$, $\overline{q}$ goes from  small-$\overline{q}$ (which vanishes in the limit of large-$N$) to $\overline{q}\sim 1$. Note that, for large $\mu$, the alleles $\pm 1$ have roughly equal probability  (namely, $h_s(j)\sim0.5$ for each locus), and hence the steady state does not significantly differ from the initial one. On the contrary, for small  $\mu$,  the system reaches fixation (i.e., $h_s(j)\sim0$ for each locus). In this case, the individuals are just clones, namely, for each realization, the population is represented by a unique genome that is a generic combination of $\pm 1$ in a neighborhood of the phenotype optimum (exactly in the optimum for $\omega \rightarrow \infty$). The crossover from the state with small $\overline{q}$ (large heterozygosity) to the state with $\overline{q}\sim 1$  (small heterozygosity) is  characterized by a maximum in the susceptibility, $\overline{\chi}$. The value of $\mu$ corresponding to such a maximum is a monotonic increasing function of $\omega$. Moreover, concerning its dependence on $M$, it is interesting to observe that it simply scales as $1/M$, as one can expect since $2 \mu M$, representing the mutation rate per individual, is the relevant quantity, ruling the mutations during evolution.  From Fig. \ref{fig:1}, it can be easily observed that for weak selection strength, roughly $\omega < 0.4$ (with blue circles in figure corresponding to $\omega =0.4$), one has a smooth crossover that becomes abrupt for larger $\omega$.

To better analyze the nature of these steady states, and the crossover from small-$\overline{q}$ states to large-$\overline{q}$  ones, we perform the following numerical experiment.  For any  value of $\omega$, starting from a configuration at high mutation rate, we decrease $\mu$ at a given rate $\dot{\mu}\equiv \Delta\mu/\Delta n$ ($n$ denoting the generation number). In other words, the system is kept at a given value of the mutation rate for an interval $\Delta n$, and, at the end of it, $\overline{q}$ and $\overline{\chi}$ are measured. Afterward, the value of $\mu$ is decreased of $\Delta\mu$ and the procedure is iterated till $\mu$ reaches zero. At this point the procedure is inverted and $\mu$ is increased at the same rate, in analogy to a physical system, first cooled and then heated at given rate. As usual in thermodynamics, for any $\mu$, two
states are considered macroscopically equivalent if the measured values of
$\overline{q}$ and $\overline{\chi}$ coincide.
\begin{figure}[ht]
\begin{center}
\includegraphics[width=6cm]{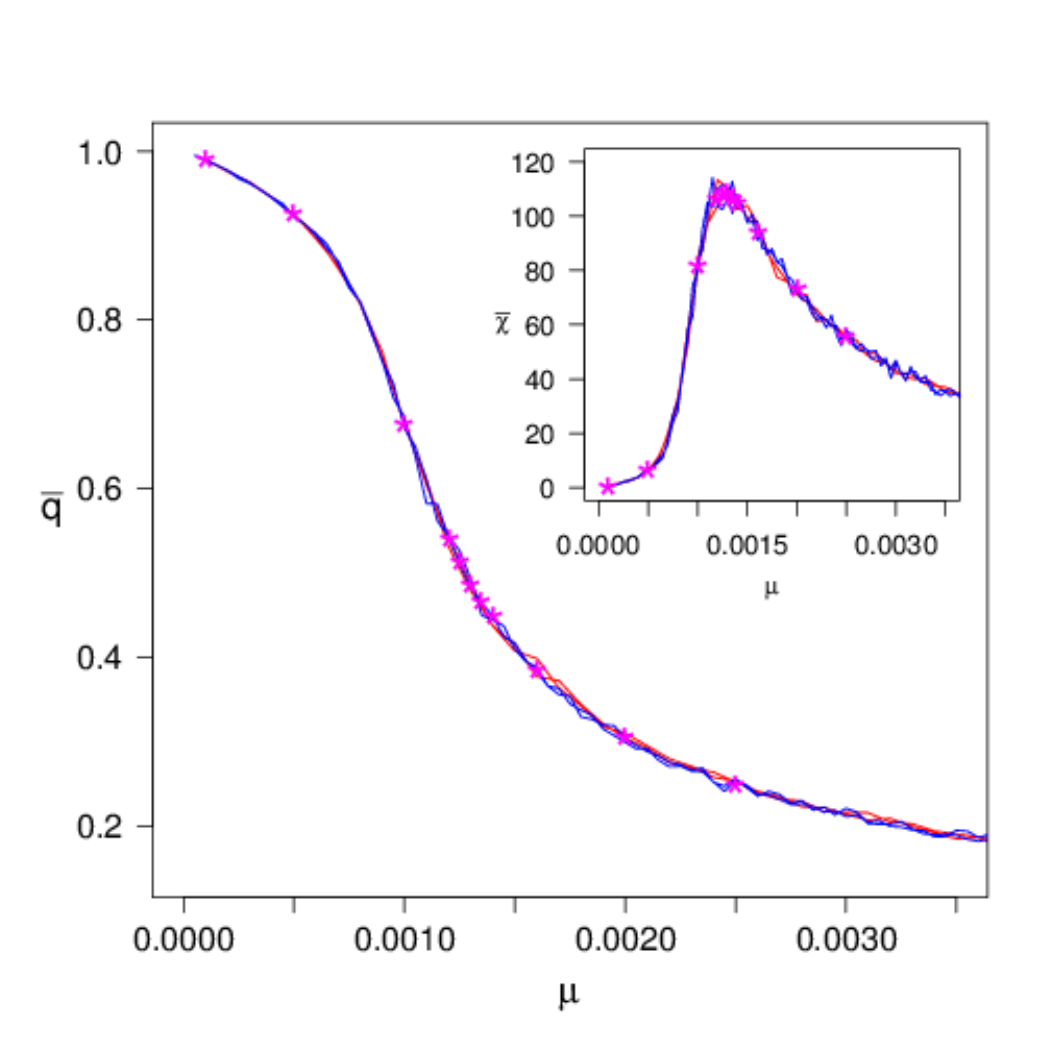}
\end{center}
\caption{
{\bf Main frame}:
Order parameter, $\overline{q}$ vs $\mu$, for $ \omega=0.1$ and $N=1000$. The full lines are obtained first cooling the system at fixed $\dot{\mu}$ (the red line corresponds to $10^{-8}$ and the blue line to $10^{-9}$), and then heating it at the same rate. The pink stars correspond to the steady states. {\bf Inset}: Susceptibility, $\overline{\chi}$ vs $\mu$, with the same symbols as in the main frame.}
\label{fig:2}
\end{figure}
As we will show in the following sections, this procedure confirms the presence of two different regimes, for $\omega<\omega_c$ and $\omega \geq \omega_c$, respectively, where the threshold $\omega_c\simeq 0.4 + {\cal O}(N^{-1})$.

\subsection{Small selective pressure}
In Fig. \ref{fig:2}, $\overline{q}$ and $\overline{\chi}$ vs $\mu$ are plotted
for $\omega=0.1$ at two different values of the cooling rate.
As we see in figure, for small enough cooling rate, the curves do not depend
on the {\em cooling} rate, $\dot{\mu}$, and the two branches, obtained by
decreasing and increasing $\mu$ respectively, always coincide.
These behaviors are observed for each value of $\omega<\omega_c$ (small
selective pressure).
Moreover, the states obtained with this procedure in the limit of small
cooling rate coincide with the above defined steady states (pink stars in Fig.
\ref{fig:2}), and in some sense, these states can be considered
{\em equilibrium} states of the system.
The crossover here observed from large to small heterozygosity, ruled by the
mutation rate and characterized by a maximum in the susceptibility, strongly
resembles a continuous (second order) transition in a physical system.
This observation suggests to study this transition as a usual critical
phenomenon.

From the intersection of the fourth order cumulant, evaluated for different sizes of the system, the critical mutation rate, $\mu_c \equiv \lim_{N \rightarrow \infty} \mu_c(N)$, is estimated, and a finite size scaling analysis is performed  in order to evaluate the critical exponents of the transition.

For an Ising model with vanishing magnetic field \footnote{For vanishing magnetic field ($H=0$) the Ising model undergoes  a second order phase transition from a disordered paramagnetic phase (vanishing magnetization) to an ordered ferromagnetic one (not vanishing magnetization), at temperature $T_c$.  In the ferromagnetic phase ($T\le T_c$), the magnetization $\rightarrow 0$ at the critical temperature as a power law with exponent $\beta$. For fixed temperature $T<T_c$, a first order transition controlled by the magnetic field, is found for vanishing $H$.
}, the reduced fourth order cumulant of the order parameter \cite{binder} is given by
\begin{equation}
U_4=1- \frac{\langle m^4\rangle}{3\langle m^2\rangle^2},
\label{u4}
\end{equation}
where $m$ is the magnetization. Following Ref. \cite{binder}, as the system size $N\rightarrow \infty$, $U_4\rightarrow 0$ for $T > T_c$ and $U_4 \rightarrow 2/3$ for $T < T_c$ .  For large enough values of the size $N$, all curves representing $U_4$ as a function of temperature cross in a point whose location gives the critical point.

A natural extension of Eq. (\ref{u4}) to our biological system is
\begin{equation}
\overline{U}_4\equiv\frac{1}{M}\sum_{j=1}^M U_4(j),
\end{equation}
where
\begin{equation}
U_4(j)=1- \frac{\langle q(j)^4\rangle}{3\langle q(j)^2\rangle^2}
\end{equation}
is the fourth order cumulant of $j$-th locus, and $\overline{U}_4$ denotes the average over the $M$ loci.
\begin{figure}[ht]
\begin{center}
\includegraphics[width=6cm]{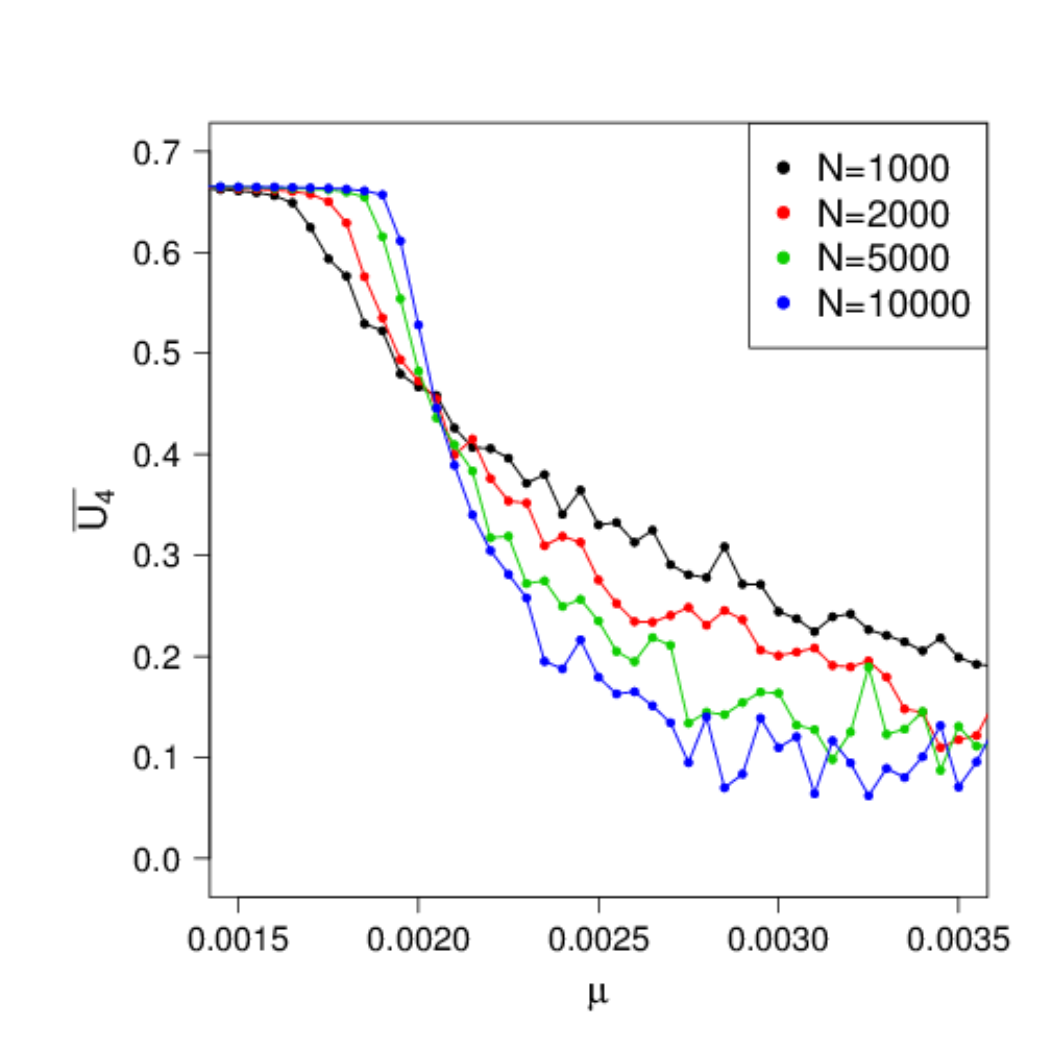}
\end{center}
\caption{
$\overline{U}_4$ vs $\mu$, for $\omega=0.2$ and different values of $N$. The crossing point is $\mu_c\sim 0.0021$. The continuous lines are guides for eyes.}
\label{fig:3}
\end{figure}

\begin{figure}[ht]
\begin{center}
\includegraphics[width=6cm]{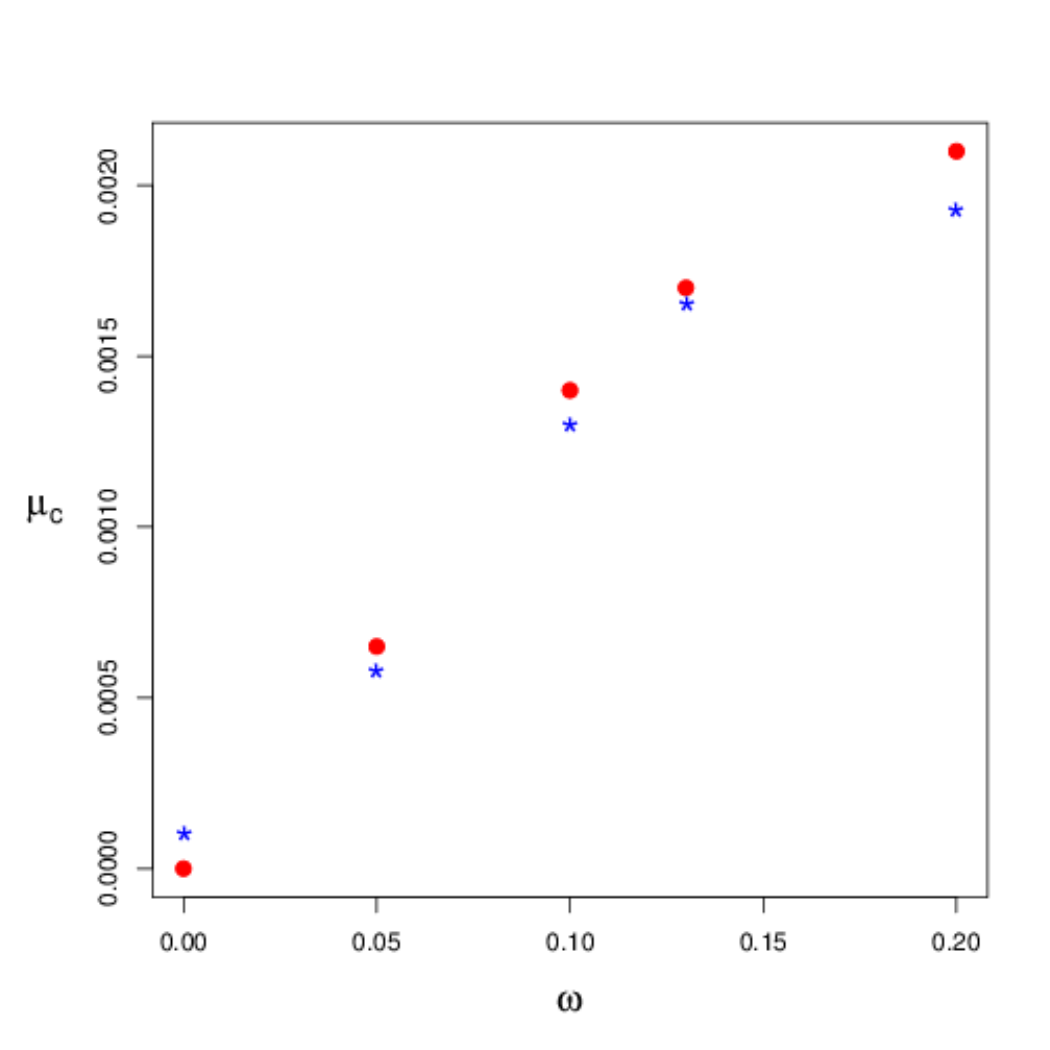}
\end{center}
\caption{Critical mutation rate, $\mu_c$ (red circles) vs $\omega$, compared with  $\mu_c(N)$, for $N=1000$ (blue stars).}
\label{fig:4}
\end{figure}



In the present case, increasing the mutation rate,  in each locus $j$  it is observed a transition from a disordered phase to an ordered one. This is in perfect analogy with the Ising model. Hence, we expect that varying the size $N$ of the system, all curves for $\overline{U}_4$ as a function of $\mu$ cross in a point which provides the critical mutation rate $\mu_c \equiv \lim_{N \rightarrow \infty} \mu_c(N)$. Note that $\mu_c$ keeps a dependence on $\omega$.

In Fig. \ref{fig:3}, $\overline{U}_4$ is plotted as a function of $\mu$ for different values of the size $N$, having fixed $\omega=0.2$. As expected, in the limit of large $N$ we find that $\overline{U}_4$ tends to $2/3$ and to zero for small and large mutation rate, respectively. The crossing point, $\mu_c$, represented by the red circles in Fig. \ref{fig:4}, increases by increasing $\omega$ and vanishes in the limit $\omega\rightarrow 0$. Blue stars in the same figure correspond to the maximum points of the susceptibility, namely $\mu_c(N)$ for $N=1000$. We find that $\mu_c(N)$ is smaller than $\mu_c$ for each non vanishing $\omega$, whereas this behavior reverses for $\omega=0$.

Next, the critical behavior of the order parameter and of the susceptibility are studied. Extending the predictions from finite size scaling analysis in the Ising model \cite{binder} to the present system, we expect that near the critical point
\begin{eqnarray}
\overline{q}&=&N^{-a} q_0\left(\epsilon N^c\right),\nonumber \\
\overline{\chi}&=& N^b \chi_0\left(\epsilon N^c\right),
\label{eq:scaling}
\end{eqnarray}
where $\epsilon=(\mu-\mu_c)/\mu_c$, and $q_0$ and $\chi_0$ are scaling functions. 

A finite size scaling analysis allows to evaluate the exponents $a$, $b$, and $c$. 
For each value of $\omega$,  
$N^a\overline{q}$ and $N^{-b}\overline{\chi}$ are plotted 
as a function of  $\epsilon N^c$, where $a$, $b$, and $c$  are chosen in order to rescale the curves  for different $N$ onto a unique one (data not shown).

In a $d$-dimensional physical system, $a$, $b$ and $c$  are related to the critical exponents, $\nu$, $\beta$ and $\gamma$, by the following relations \cite{binder}: $a=\beta/\nu d$, $b=\gamma/\nu d$, and $c=1/\nu d$, where  $\nu$, $\beta$ and $\gamma$ depend on the euclidean dimension, and tend to the mean field exponents in the limit of high dimension $d$. Although the space dimension is here not defined at all, we can evaluate $\beta$ and $\gamma$, from $a$, $b$ and $c$, as $\beta=a/c$ and $\gamma=b/c$. In Table \ref{tabella1}, $\mu_c$ and the scaling exponents obtained for different $\omega$ are listed \footnote{Note that the scaling relation $2\beta+\gamma=\nu d$, which in terms of $a$ and $b$ becomes $2a+b=1$, is almost everywhere verified.}.
Although the errors are rather large (of the order of the $10$\%) and further analysis is necessary to confirm these findings, the critical exponents seem to change along the critical line, and to tend to the mean field Ising critical exponents (i.e., $\beta=0.5$ and $\gamma=1$) by approaching $\omega_c$. This result can be interpreted in the following way.
The correspondence between micro-states with optimum phenotype and energy minima 
in physical systems is rather natural.
Following this analogy, the growth of $\omega$ would correspond to increase
the barriers between two minima. This suggests the possibility that the
selective pressure in some sense plays the role of the euclidean dimension,
controlling the {\it energy} landscape of the system,
and the maximum selective pressure corresponds to the mean field limit, where
energy barriers between different minima become infinite.

For $\omega=0$, no crossing point is observed in the fourth order cumulant.
Consistently, the maximum point of the susceptibility $\mu_c(N)$ goes to zero as $1/N$ 
(data not shown)
in the limit $N\rightarrow\infty$, and
the following trivial finite size scaling is found (data not shown):
\begin{eqnarray}
\overline{q}&=&q_0\left(\mu N\right),\nonumber\\
\overline{\chi}&=& N \chi_0\left(\mu N\right).
\end{eqnarray}
It is worth observing, that due to this scaling behavior of $\overline{q}$ and $\overline{\chi}$, the heterozygosity, $\overline{H}_s$, defined in Eq. (\ref{HS}) results scale free for $\omega=0$. Moreover, in absence of selective pressure, since the loci are independent, there is no dependence on $M$ at all.
\begin{table}[!ht]
\centering
\begin{tabular}{|c|c|c|c|c|c|c|}
\hline
$\omega$ & $\mu_c$ & $a $  & $b$  & $c$ & $\beta$  & $\gamma$\\
  \hline
  $0$ & $0$ & $0$ & $1$ & $1$ & $0$ & $1$
\\
  $0.05$ & $0.00065$ & $0.15$
  & $0.7$ & $0.5$ & $0.3$ & $1.4$
 \\
  $0.1$ & $0.0014$ & $0.175$
  & $0.65$ & $0.5$ & $0.35$ & $1.3$
\\
  $0.13$ & $0.0017$ & $0.2$
  & $0.6$ & $0.5$ & $0.4$ & $1.2$
\\
  $0.2$ & $0.0021$ & $0.25$
  & $0.6$ & $0.5$  & $0.5$ & $1.2$
\\
  \hline
\end{tabular}
\caption{Critical mutation rate and scaling exponents, for different $\omega$.}
\label{tabella1}
 \end{table}
\begin{figure}[ht]
\begin{center}
\includegraphics[width=6cm]{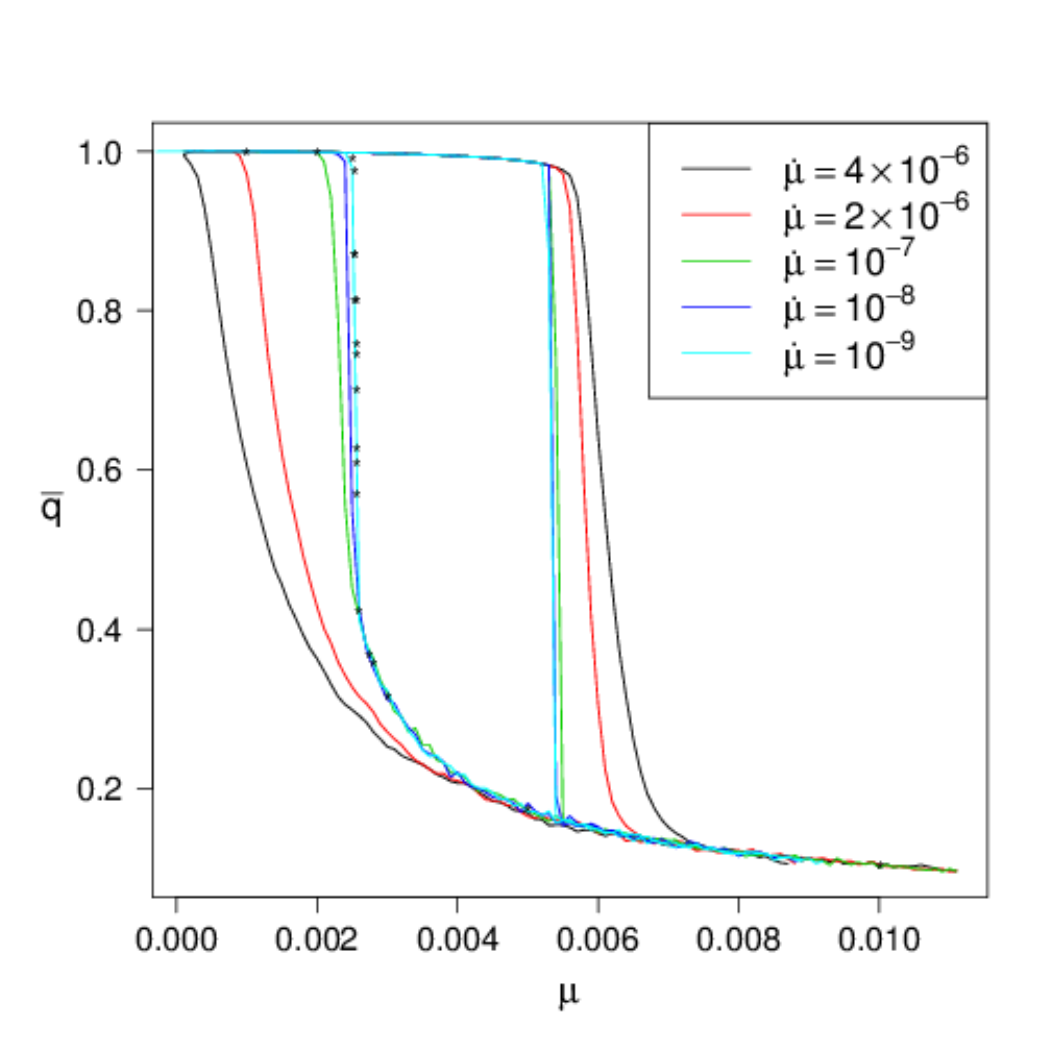}
\end{center}
\caption{
Order parameter, $\overline{q}$ vs $\mu$, for $N=~1000$ and $\omega \rightarrow\infty$. The different cycles have been obtained by using different cooling rates.The black stars reproduce data plotted in Fig. \ref{fig:1} for $\omega\rightarrow\infty$, obtained by following the system up to  $2\cdot 10^6$ generation numbers.
}
\label{fig:8}
\end{figure}

\subsection{Hysteresis cycles  at large selective pressure}
Interestingly, in the region of large selective pressure, for $\omega\ge\omega_c$,  a metastable hysteresis appears between small and close to $1$ values of $\overline{q}$, as shown in Fig. \ref{fig:8}.  Again, for large mutation rate, the system is at equilibrium in states with small-$\overline{q}$, and, for small mutation rate, is at equilibrium in states with $\overline{q}\sim 1$. However, the two branches, at small$-\overline{q}$ and $\overline{q}\sim 1$ respectively, are both observed for intermediate values of the mutation rate, depending on the pattern of $\mu$-variation.  Although by decreasing the cooling rate the hysteresis cycle shrinks, we always see two well distinct branches on our observation time scales. This behavior is reminiscent of a discontinuous (first order) transition,  where metastable hysteresis is usually observed. In this case, the distinction of long-lived metastable states from equilibrium states is rather difficult, since the lifetime of the metastable states may be longer than the observation time. As it can be seen in Fig. \ref{fig:8}, the states, obtained decreasing $\mu$ at small $\dot{\mu}$, coincide with the steady states, reached by the system for very large generation numbers. Hence, in this case the so-called steady states, which are stationary  on our observation time scales, are likely metastable. Note that in Fig. \ref{fig:8} the hysteresis curves are plotted for $\omega \rightarrow \infty$, where this phenomenon is more evident.

\begin{figure}[ht]
\begin{center}
\includegraphics[width=6cm]{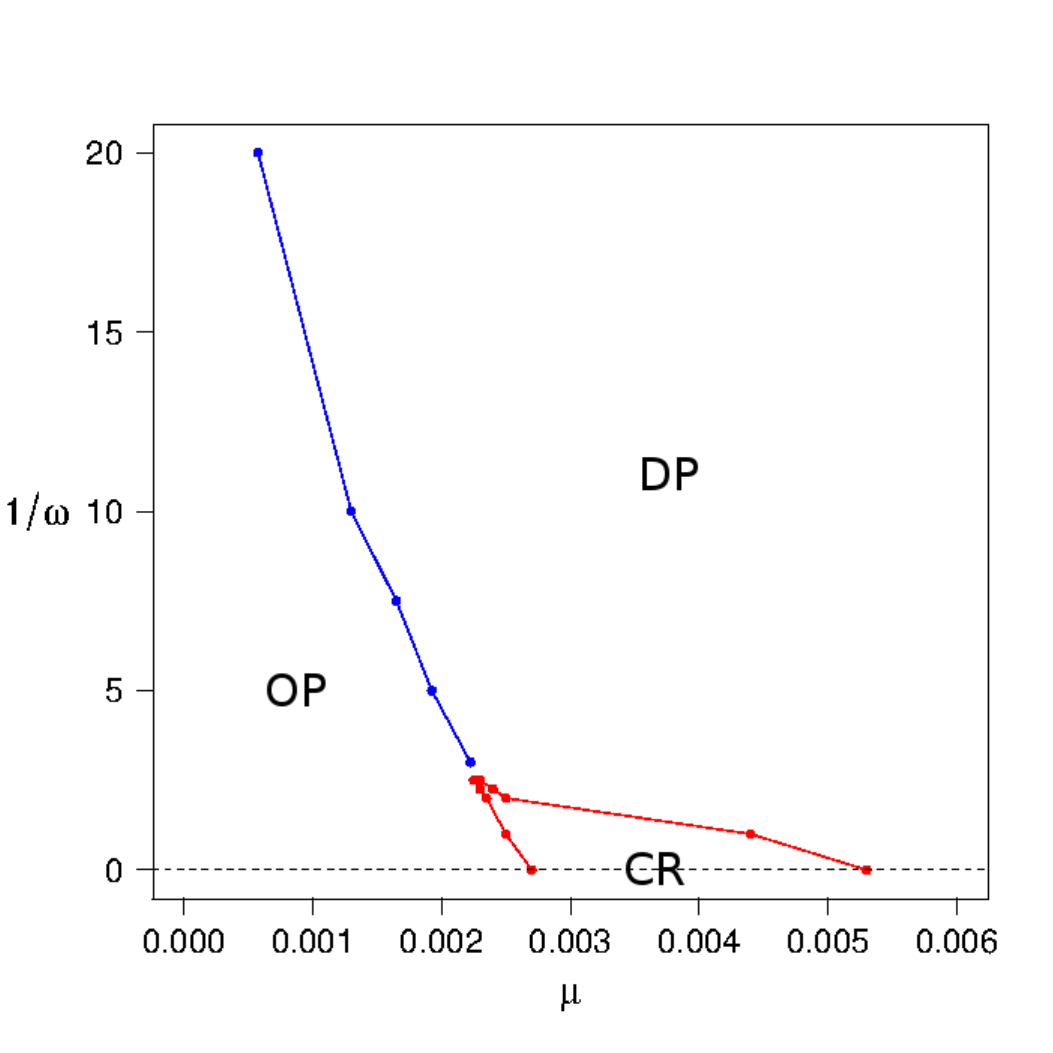}
\end{center}
\caption{Phase diagram in the plane ($\mu$, $1/\omega$) for a system of size $N=1000$ (see text for explanations). DS indicates the Disordered Phase, OS the Ordered Phase and CR the Coexistence Region. The continuous lines are guides for eyes. }
\label{fig:9}
\end{figure}

Our findings are efficaciously summarized in Fig. \ref{fig:9}, where the phase diagram for a system of $N=1000$ individuals is shown in the plane  ($\mu$, $1/\omega$).  For $\omega<\omega_c$, we plot $1/\omega$ as function of the maximum point of $\overline{\chi}$, $\mu_c(\omega,N)$ (data already shown in Fig.
\ref{fig:4}). The blue line should give, in the {\em thermodynamic} limit, a line of critical points, where the continuous transition from small-$\overline{q}$ to $\overline{q}=1$ phase should be observed. In the region at large $\mu$, the system is found in the Disordered Phase (DP), and in the region at small $\mu$ it is found in the Ordered Phase (OP). Above $\omega_c$, data depend on the cooling rate and the susceptibility displays two maxima, depending on the pattern of $\mu$-variation.  Red circles in Fig. \ref{fig:9} correspond to the maximum points of $\overline{\chi}$, along the hysteresis loop obtained at the smallest cooling rate, $\dot{\mu}=10^{-9}$. In this region, the system behaves as a physical system undergoing a discontinuous transition controlled by the mutation rate. Between the two red lines, the two phases coexist.

\subsection{Heterozygosities}

\begin{figure}[ht]
\begin{center}
\includegraphics[width=6cm]{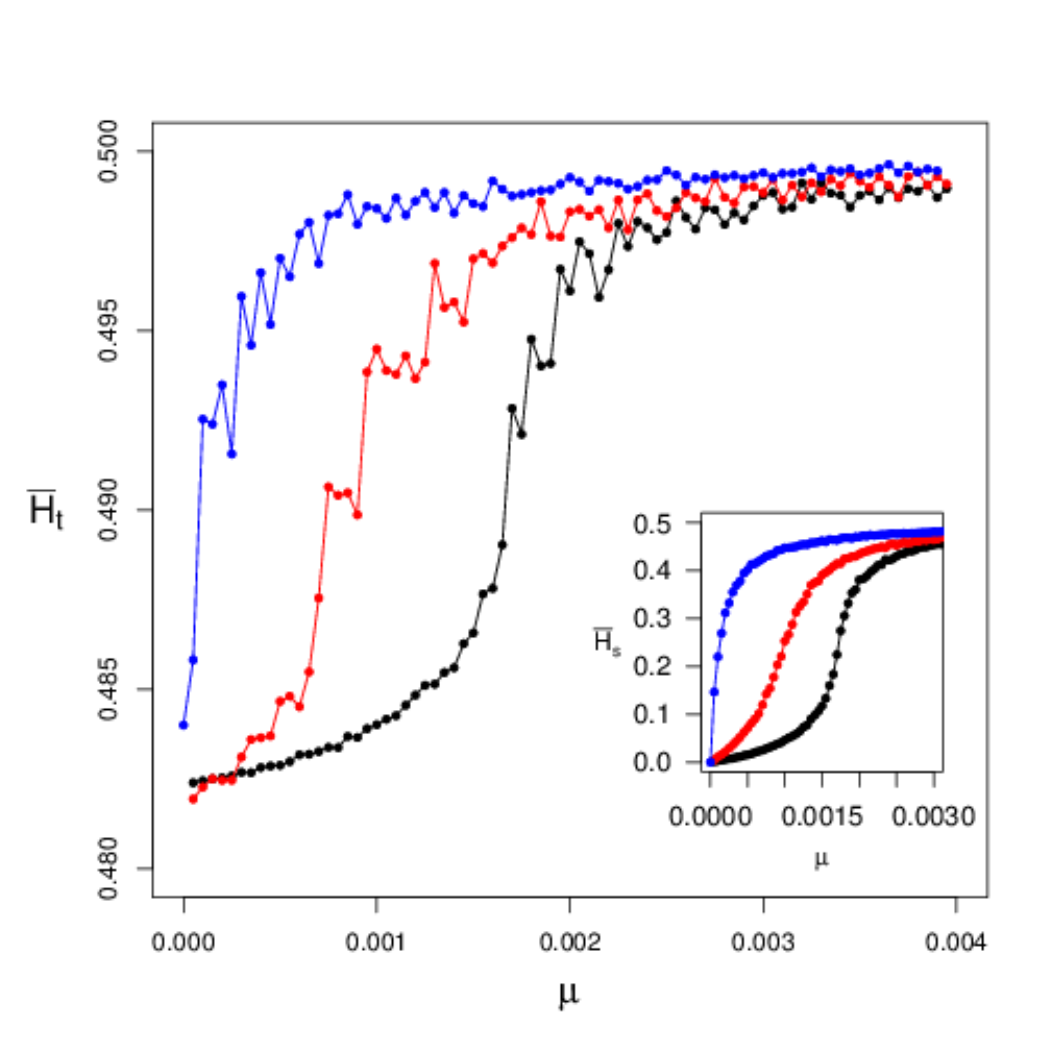}
\end{center}
\caption{
{\bf Main frame}: $\overline{H}_t$ vs $\mu$, in the steady states for $N=1000$ and $\omega=0,~0.1,~0.2$ (from left to right).  {\bf Inset}: $\overline{H}_s$ vs $\mu$, in the steady states  for $N=1000$ and $\omega=0,~0.1,~0.2$ (from left to right).  The continuous lines are guides for eyes.  }
\label{fig:10}
\end{figure}

The analysis on the dependence of the order parameter, $\bar{q}$, on the mutation rate, $\mu$, is also carried out for the expected heterozygosity of the single population, $\overline{H}_s$, and for the expected heterozygosity measured on the set of populations as a whole,  $\overline{H}_t$.
As for $\overline{q}$, hysteresis cycles are observed for $\omega\ge\omega_c$ (data not shown).
Interestingly, $\overline{H}_s$ and $\overline{H}_t$ display different behaviors for small mutation rate.  Figs.  \ref{fig:10} 
shows that $\overline{H}_s\sim 0.5$ and $\overline{H}_t\sim 0.5$, for large mutation rate, whereas $\overline{H}_s\sim 0$ and $\overline{H}_t\sim 0.5$ (we expect $\overline{H}_t = 0.5$ for $N_p \rightarrow \infty$), for small mutation rate, where the independently evolving populations (although initially identical) reach  fixation in generally different (but macroscopically equivalent) micro-states, developing a genetic diversity. This phenomenon is in some sense analogous of the spontaneously symmetry breaking in physical system.

\section{Comparison with literature and Conclusions}

In summary, we have analyzed the evolution of a population of $N$ diploid individuals, sexually reproducing with random mating, evolving under the effect of a Gaussian viability depending on an additive polygenic trait. Using the standard tools of Statistical Mechanics, we show that the system displays a complex phase diagram with a transition from a disordered to an ordered phase, controlled by the mutation rate. We provide the phase diagram in the ($\mu$, $\omega$) plane, showing that the order of the transition changes depending on the strength of selection, being continuous for  weak selective pressures and discontinuous for strong ones. Similar findings are expected for a population of $2N$ haploid individuals.

{Many analogies are found in literature between evolution and Statistical Mechanics, and they  are not all equivalent. In our picture, the mutation rate plays the role of temperature in statistical physics (and $N$ plays the same role of the finite dimension in physics systems), in agreement with Leuthausser's analogy between the Eigen model and an Ising system \cite{Leuthausser}. In other formulations (see for instance \cite{Sella2005,Mustonen}), temperature is instead related to population size. In Ref. \cite{Sella2005}, small mutation rates are considered, and populations are always in our fixation limit, i.e. they are made by clones. Then, the system state is a point in the genome space (which here is a $2M$ dimensional space), and not a point in the $2MN$ configurational space of individuals. In this limit, the mutation rate does not affect the steady state, and can merely influence the dynamics of the system. These two divergent points of view can be reconciled if one thinks about the main source of stochasticity that for large $\mu$ and $N$ is dominated
by the mutation rate (present analysis), and that on the contrary, for small $\mu$ and $N$ is dominated by the random drift (see for instance
\cite{Sella2005}). Since the quantity representing the main source of stochasticity is the natural candidate to play the role of temperature, this would
explain the different approaches present in literature.

The presence of phenomena analogous to phase transitions is also not new in biological evolution, in particular in the quasi-specie context
\cite{eigen,Leuthausser,Tarazona,Stadler,Galluccio,ParkDeem},
in strict analogy with the critical mutation rate here found,
the error threshold is the critical value of the mutation rate, below
that the population is closely centered around the fitness peak, and above that it is roughly
distributed over all the accessible space, losing the favorable sequence.
We observe that $\mu_c$  is a monotonic increasing function of $\omega$, with a fix point in $\mu_c=0$ for $\omega=0$ (obviously, in absence of selective pressure, the system is always in the disordered phase). Increasing the selective pressure, values of $\mu_c$ roughly between $6\cdot 10^{-4}$ and 
$2\cdot 10^{-3}$ are observed.
Since the relevant quantity is the mutation rate per individual, $2\mu M$, we expect that $\mu_c$ simply scales as $1/M$  (similar behaviors are
observed for the error threshold in the single-peaked landscape \cite{baakegabriel}).
Thus, we can speculate that, in viral populations, where mutation rate is estimated between $10^{-4}$ and $10^{-5}$, systems near these transitions can exist.   

For future, we intend to study the effect of a different choice for the optimum of the viability, $p_m$. It is interesting to explore how our findings
change considering a less degenerate case (the case here considered $p_m=0$ is the most degenerate one), or even a
non-reachable optimum value. In particular, we intend to investigate how the hysteresis cycles found at large selective pressure depend on this
particular choice. Preliminary simulations show that the model with a different optimum ($p_m \ne 0$) displays qualitatively similar behavior, with $\omega_c$
decreasing as $|p_m|$ increases. However, further work is necessary to confirm this behavior and to understand its meaning.
Finally, the effect of a different form for the viability will be also investigated. In particular, the model can be easily extended to
include multiple optimal phenotypes and, thus, be used to study speciation.

{\bf Acknowledgments}

The authors would like to thank L. Peliti and A. Coniglio for valuable discussions.


{\bf Author's contributions}

AF conceived the model, implemented the software and drafted the
manuscript; SC and GM conceived the study, participated in its design
and coordination, and helped to draft the manuscript. AM and GS contributed
to the discussion of the results, and helped to draft the manuscript.
All authors have read and approved the final manuscript.

\end{document}